\documentstyle[12pt]{article}
\newcommand{\resection}[1]{\setcounter{equation}{0}\section{#1}}

\def\MPL #1 #2 #3 {{\sl Mod.~Phys.~Lett.}~{\bf#1} (#3) #2}
\def\NPB #1 #2 #3 {{\sl Nucl.~Phys.}~{\bf B#1} (#3) #2}
\def\PLB #1 #2 #3 {{\sl Phys.~Lett.}~{\bf B#1} (#3) #2}
\def\PR #1 #2 #3 {{\sl Phys.~Rep.}~{\bf#1} (#3) #2}
\def\PRD #1 #2 #3 {{\sl Phys.~Rev.}~{\bf D#1} (#3) #2}
\def\PRL #1 #2 #3 {{\sl Phys.~Rev.~Lett.}~{\bf#1} (#3) #2}
\def\RMP #1 #2 #3 {{\sl Rev.~Mod.~Phys.}~{\bf#1} (#3) #2}
\def\ZPC #1 #2 #3 {{\sl Z.~Phys.}~{\bf C#1} (#3) #2}
\def\IJMP #1 #2 #3 {{\sl Int.~J.~Mod.~Phys.}~{\bf#1} (#3) #2}
\def\Zint{{Z \kern -.45 em Z}}
\def\complex{{\kern .1em {\raise .47ex \hbox
{$\scriptscriptstyle |$}}
\kern -.4em {\rm C}}}
\def\real{{\vrule height 1.6ex width 0.05em depth 0ex
\kern -0.06em {\rm R}}}

\newcommand{\be}{\begin{equation}}
\newcommand{\ee}{\end{equation}}
\newcommand{\bea}{\begin{eqnarray}}
\newcommand{\eea}{\end{eqnarray}}
\newcommand{\nn}{\nonumber}

\newcommand{\AAA}{{\cal A}}

\newcommand{\GG}{{\cal G}}
\newcommand{\LL}{{\cal L}}
\newcommand{\MM}{{\cal M}}

\def\bra#1{\langle #1 |}
\def\ket#1{| #1\rangle }
\begin{document}

\vspace*{1cm}
\begin{center}
  \begin{Large}
  \begin{bf}
Algebras, Derivations and Integrals\\
  \end{bf}
  \end{Large}
\end{center}
  \vspace{5mm}
\begin{center}
  \begin{large}
R. Casalbuoni\footnote{On leave from
Dipartimento di Fisica, Universit\`a di Firenze,
I-50125 Firenze, Italia}\\
  \end{large}
D\'epartement de
Physique Th\'eorique, Universit\'e de Gen\`eve\\ CH-1211 Gen\`eve
4, Suisse \\ {\tt{e-mail: CASALBUONI@FI.INFN.IT}}\\
\end{center}
\vspace{0.2cm}
\begin{center}
  \begin{bf}
  ABSTRACT
  \end{bf}
\end{center}
  \vspace{0.2cm}

\noindent
In the context of the integration over algebras  introduced
in a previous paper, we obtain several results for a
particular class of associative algebras with identity. The
algebras of this class are called self-conjugated, and they
include, for instance, the paragrassmann algebras of order
$p$, the quaternionic algebra and the toroidal algebras. We
study the relation between derivations and integration,
proving a generalization of the standard result for the
Riemann integral about the translational invariance of the
measure and the vanishing of the integral of a total
derivative (for convenient boundary conditions). We consider
also the possibility, given the integration over an algebra,
to define from it the integral over a subalgebra, in a way
similar to the usual integration over manifolds. That is
projecting out the submanifold in the integration measure.
We prove that this is possible for paragrassmann algebras of
order $p$, once we consider them as subalgebras of the
algebra of the $(p+1)\times(p+1)$ matrices. We find also
that the integration over the subalgebra coincides with the
integral defined in the direct way. As a by-product we can
define the integration over a one-dimensional Grassmann
algebra as a trace over $2\times 2$ matrices.

\vspace{.2 cm}
\begin{center}
UGVA-DPT 1998/03-1000
\end{center}
\vspace{.5 cm}
\newpage


\newpage
\resection{Introduction}

Quantum mechanics has modified in a  very profound way the
classical understanding of the phase space of a physical
system, making it non-commutative. This is reflected in a
drastical change of the mathematics involved by promoting
the classical phase space variables to operators acting on a
Hilbert space. However, the flavor of the classical
description is retained in the path-integral formulation of
quantum mechanics. In this case, although in the context of
a rather different physical interpretation, one retains the
concept of trajectories in the phase space. The situation
changed again with the discovery of supersymmetric theories
\cite{SUSY}. Although there are no
experimental hints about their physical reality, the beauty
of the mathematical structure involved has lead to an
enormous amount of efforts in their understanding. But then,
being supersymmetry tied to space-time invariance, one has
to give up also to path-integration in terms of commuting
variables. The space where the Feynman trajectories are
defined gets enlarged to involve anticommuting variables
(elements of a Grassmann algebra) related in an unavoidable
way to the phase space coordinates by supersymmetry
transformations. Also, considering the simple example of the
supersymmetric particle \cite{casalbuoni}, one realizes that
due to the constraints involved,  the space-time variables,
$x_\mu$, loose their commutation properties. This aspect is
not often emphasized, because it can be avoided by requiring
the constraints to be satisfied as conditions on the
physical states (the definition of chiral superfields),
rather than solve them directly \cite{casalbuoni}. Finally
the matrix realization of the $M$-theory introduces
non-commuting coordinates (matrix-valued) for the
$D0$-branes \cite{BFSS}. Following these considerations we
found interesting to introduce, for a general algebra, the
concept of integration \cite{integrale}, since it would play
a vital role in the definition of the path-integration for
these more general theories. To this end we started from the
general approach to noncommutative geometry
\cite{connes,drinfeld}. That is from the observation that,
in the commutative case, a space can be reconstructed from
the algebra of its functions. Starting directly with the
algebraic structure one can face situations where there are no
concrete realizations of the space. Said that we have
still to find a way to define the integration over an
algebra. This can be done by lifting up to the algebra level
the concept of integration over the space. To this end, let us
look at the physics beyond the path-integral formalism.
The physical amplitudes satisfy the composition law in an
automatic way within this formalism, and this arises from
the completeness relation which, in the case of a one-dimensional
system, reads
\be
\int\ket x\bra x\; dx=1
\ee
Suppose that we have a set of orthonormal  states in our
Hilbert space, $\{\ket{\psi_n}\}$. Then we can convert the
completeness relation in the space $\real^1$ into the
orthogonality relation for the wave functions
$\psi_n(x)=\langle x|{\psi_n}\rangle$,
\be
\int\langle\psi_m |x\rangle\langle x|\psi_n\rangle
dx=\int\psi_m^*(x)\psi_n(x)=\delta_{mn}
\label{fundamental}
\ee
On the other side, given this equation, and the
completeness relation for the set $\{\ket{\psi_n}\}$, we can
reconstruct the completeness in the original space
$\real^1$, that is the integration over the line. Now, we
can translate the previous properties of the set
$\{\ket{\psi_n}\}$, in the following two statements
\begin{enumerate}
\item The set of functions $\{\psi_n(x)\}$ span a vector space.
\item The product $\psi_n(x)\psi_m(x)$ can be expressed as a linear
combination of the functions $\psi_n(x)$, since the set
$\{\psi_n(x)\}$ is complete.
\end{enumerate}
All this amounts to say that the set $\{\psi_n(x)\}$ is a
basis  of an algebra. In order to capture completely the
context of eq. (\ref{fundamental}), we need also to
understand the general meaning of $\psi_n^*(x)$. From the
completeness it follows that $\psi_n^*(x)$ can be expressed
as a linear combination of the functions $\psi_n(x)$,
\be
\psi_n^*(x)=\sum_m\psi_m(x)C_{mn}
\ee
The matrix $C$ has to satisfy certain conditions that  we
will discuss in the text. In the following we will consider
associative algebras with identity and with a matrix $C$
satisfying suitable conditions. These algebras will be
called self-conjugated. In these cases we will define the
integral over the algebra by eq. (\ref{fundamental})
\be
\int_{(x)}\sum_j x_j C_{ji} x_k=\delta_{ik}
\label{1.4}
\ee
where $\{x_i\}$ is a basis of the algebra. The properties
that the $C$ matrix has to satisfy are such that the
integral of an arbitrary element of the algebra, $\int_{(x)}
x_i$, must be compatible with (\ref{1.4}) and with the algebra product.
This will be
discussed in the following Section. This procedure and his
motivations have been widely illustrated, in many examples,
in ref. \cite{integrale}. There we discussed also how to deal
with important cases as the bosonic oscillator, or the
$q$-bosonic oscillator algebras, where a suitable $C$
matrix does not exist. In this paper we will be interested
in discussing some general result valid for self-conjugated
algebras. In more detail we will prove the following
results:
\begin{enumerate}
\item A theorem relating derivations on the algebra
satisfying the integration by part rule (the vanishing of
the integral of the derivation of an arbitrary element of
the algebra) and automorphisms leaving invariant the
integration measure. This is an extension of the  theorem
relating the invariance of the Riemann integral under
translations and the vanishing of a total derivative (for
convenient boundary conditions). We stress this point
for its relevance within the
path-integral
approach, where the validity of the Schwinger's quantum
principle depends precisely on the validity of this theorem.
\item All  inner derivations, that is the derivations given by
commutators, satisfy the integration by part rule.
\item The algebra of the $N\times N$ matrices, $\AAA_N$,
is a self-conjugated algebra, with the integration given by
the trace. The integration by part rule corresponds here to
the cyclic property of the trace.
\item  Given the integral over an algebra one
can think of inducing it on a subalgebra. This is done in the
particular  case of a paragrassmann algebra of order $p$,
$\GG_p$, (that is generated by an element $\theta$ such that
$\theta^{p+1}=0$) thought as a subalgebra of
$\AAA_{p+1}$, the algebra of the $(p+1)\times(p+1)$ matrices. The idea is to
project out the
subalgebra from the algebra, as one defines the integration
over a submanifold by projecting it out from the manifold.
In fact we will express the corresponding integral as a
trace of the representative of the elements of $\GG_p$ in
$\AAA_{p+1}$, times an operator which projects out $\GG_p$
from $\AAA_{p+1}$. In particular, this will allow us to get
the integral over a Grassmann algebra ($p=1$) as a trace
over $2\times 2$ matrices.
\end{enumerate}

This paper is organized as follows: in Section  2 we will
recall the main concepts necessary to define the integration
over an algebra \cite{integrale}. In Section 3 we will study
the relation between self-conjugated and involutive
algebras. In Section 4 we will define the integration over
the algebra of the $N\times N$ matrices. In Section 5 we
will introduce the concept of derivation and we will derive
the results 1), 2) and 3) mentioned above. The result 4) will
be obtained in Section 6.

\resection{Algebraic integration}

We recall here some of the concepts introduced in
\cite{integrale}, in order to define the integration
rules over a generic algebra. We start by considering
an algebra $\AAA$ given by $n+1$ basis elements $x_i$, with
$i=0,1,\cdots n$ (we do not exclude the possibility of
$n\to\infty$, or of a continuous index). We assume the
multiplication rules
\be
x_i x_j=f_{ijk}x_k
\ee
with the usual convention of sum over the repeated
indices. For the future manipulations it is convenient to organize
the basis elements $x_i$ of the algebra in a ket
\be
| x\rangle=\left(\matrix{x_0\cr x_1\cr\cr \cdot\cr\cdot\cr
x_n\cr}\right)
\ee
or in the corresponding bra
\be
\bra x=\left(\matrix{x_0 & x_1 & \cdots & x_n}\right)
\ee
Important tools for the study of a generic  algebra are the
{\bf right and left multiplication algebras}. We define the
associated matrices by
\be
R_i|x\rangle=|x\rangle x_i,~~~~\bra {
x}L_i=x_i\bra{ x}
\label{eigenequation}
\ee
For a generic element $a=\sum_ia_ix_i$ of the algebra we
have $R_a=\sum_ia_iR_i$, and a similar equation for the left
multiplication. In the following we will use also
\be
L_i^T\ket x=x_i\ket x
\ee
The matrix
elements of $R_i$ and $L_i$  are obtained from their definition
\be
(R_i)_{jk}=f_{jik},~~~~(L_i)_{jk}=f_{ikj}
\label{matrici}
\ee
The algebra is completely characterized by the
structure constants. The matrices $R_i$ and $L_i$ are
just a convenient  way of encoding their properties.
In the following we will be interested in associative
algebras. By using the associativity condition
\be
x_i(x_jx_k)=(x_ix_j)x_k
\ee
one can easily show the following relations (all
equivalent to the previous relation)
\be
R_iR_j=f_{ijk}R_k,~~~L_iL_j=f_{ijk}L_k,~~~[R_i,L_j^T]=0
\label{associativity}
\ee
The first two say that  $R_i$ and  $L_i$ are linear
representations of the algebra, called the regular representations.
The third that the right and
left multiplications commute for associative algebras.
In this paper we will be interested in associative algebras
with identity, and such that there exists a matrix $C$,
satisfying
\be
L_i=CR_iC^{-1}
\label{self-conjugated}
\ee
\be
C^T=C
\ee
We will call these algebras self-conjugated. The condition
(\ref{self-conjugated}) is consistent with $L_i$ and $R_i$
satisfying the same algebra (see eq. (\ref{associativity})).
Therefore, the non existence of the matrix $C$ boils down
to the possibility that the algebra admits inequivalent
regular representations. This happens, for instance, in the
case of the bosonic algebra \cite{integrale}. The condition
of symmetry on $C$ can be interpreted in  terms of the
opposite algebra $\AAA^D$, defined by
\be
x_i^D x_j^D=f_{jik}x_k^D
\ee
The left and right multiplication in the dual algebra are
related  to those in $\AAA$ by
\be
R_i^D=L_i^T,~~~~L_i^D=R_i^T
\ee
Therefore the matrices $L_i^T$ are a representation of the
dual algebra
\be
L_i^TL_j^T\ket x=x_jx_i\ket x=f_{jik}L_k^T\ket x
\label{associativity2}
\ee
We see that the condition $C^T=C$ is equivalent to require
that the relation (\ref{self-conjugated}) holds also for the
right and left multiplication in the opposite algebra
\be
L_i^D=CR_i^DC^{-1}
\ee
Since we are
considering associative algebras, the requirement of
existence of an identity is not a strong one, because we can
always extend the given algebra to another associative
algebra with identity. In fact, let us call $F$ the field
over which the algebra is defined (usually $F$ is the field
of real or complex numbers). Then, the extension of $\AAA$
(call it $\AAA_1$) is defined by the pairs
\be
(\alpha,a)\in\AAA_1,~~~\alpha\in F,~~~a\in\AAA
\ee
with the product rule
\be
(\alpha,a)(\beta,b)=(\alpha\beta,\alpha a+\beta b+ab)
\ee
The identity in $\AAA_1$ is given by the pair
\be
I=(1,0)
\ee
Of course, this is the same as  adding to any element
of $\AAA$ a term proportional to the identity, that is
\be
\alpha I+ a
\ee
and defining the multiplication  by distributivity. One can
check easily that $\AAA_1$ is an associative algebra. An
extension of this type exists also for many other algebras,
but not for all. For instance, in the case of a Lie algebra
one cannot add an identity with respect to the Lie product
(since $I^2=0$). For self-conjugated algebras, $R_i$ has an
eigenbra given by
\be
\bra{xC}=\bra x C, ~~~~~\bra {xC} R_i=x_i\bra {xC}
\ee
as it follows from (\ref{self-conjugated}) and
(\ref{eigenequation}). Then, as  explained in the
Introduction, we define  the integration for a
self-conjugated algebra by the formula
\be
\int_{(x)}\ket x\bra{xC}=1
\label{2.21}
\ee
where 1 is the identity in the space of the linear mappings on the
algebra.
In components the previous definition means
\be
\int_{(x)}x_ix_k C_{kj}=f_{ikp}C_{kj}\int_{(x)}x_p=\delta_{ij}
\label{2.19}
\ee
This equation is meaningful only if it is possible to invert
it in terms of $\int_{(x)} x_p$. This is indeed the case
 if $\AAA$ is an algebra with identity (say
$x_0=I$) \cite{integrale}, because  by taking $x_i=I$ in eq.
(\ref{2.19}), we get
\be
\int_{(x)}x_j=(C^{-1})_{0j}
\label{2.20}
\ee
We see now the reason for requiring the condition
(\ref{self-conjugated}). In fact it ensures that the value
(\ref{2.20})
of the integral of an element of the basis of the algebra
gives the solution to the equation (\ref{2.19}). In fact we
have
\be
\int_{(x)}x_ix_k C_{kj}=f_{ikp}C_{kj}C^{-1}_{0p}=
(C^{-1}L_i C)_{0j}=(R_i)_{0j}= f_{0ij}=\delta_{ij}
\ee
as it follows from $x_0x_i=x_i$.
Notice that the symmetry of
$C$ allows us  to write the integration as
\be
\int_{(x)}\ket { xC}\bra x=1
\ee
which is the form we would have obtained if we had started
with  the same assumptions but with the  transposed version
of  eq. (\ref{eigenequation}). All the examples considered
in ref. \cite{integrale}, where the $C$ matrix exists, turn
out to correspond to self-conjugated algebras. The examples
we are referring to are the algebra over the circle, the
paragrassmann algebras of order $p$, and the quaternionic
 algebra. In ref. \cite{tori2} we have
considered noncommuting toroidal algebras, which also turn
out to be self-conjugated ones \cite{casalbuoni2}.

 We will define an
arbitrary function on the algebra by
\be
f(x)=\sum_if_ix_i\equiv\langle x|f\rangle
\ee
and its conjugated as
\be
f^*(x)=\sum_{ij}{\bar f}_i x_jC_{ji}=\langle f|xC\rangle
\ee
where
\be
\ket f=\left(\matrix{f_0 \cr f_1\cr \cdot\cr\cdot\cr x_n}\right),~~~~
\bra f=\left(\matrix{{\bar f}_0 & {\bar f}_1 & \cdots & 
{\bar f}_n}\right)
\ee
and ${\bar f}_i$ is the complex-conjugated  of the coefficient
$f_i$ belonging to the field $\complex$. Then a scalar
product on the algebra is given by
\be
\langle f|g\rangle =\int_{(x)}\langle f|xC\rangle
\langle x|g\rangle =
\sum_i{\bar f}_i g_i
\ee

\resection{Algebras with involution}

In  some case, as for the toroidal algebras \cite{tori2},
the matrix $C$ turns out to define a mapping which is an
involution of the algebra. Let us consider the property of
the involution on a given algebra $\AAA$. An involution is a
linear mapping $^*:{\cal A}\to{\cal A}$, such that
\be
(x^{*})^*=x,~~~~~ (xy)^*=y^*x^*,~~~~x,y\in{\cal A}
\label{definizioni}
\ee
Furthermore, if the definition field of the algebra is
$\complex$, the involution acts as the complex-conjugation
on the field itself. Given a basis $\{x_i\}$ of the algebra,
the involution can be expressed in terms of a matrix $C$
such that
\be
x_i^*=x_j C_{ji}
\ee
The eqs. (\ref{definizioni}) imply
\be
(x_i^{*})^*=x_j^*C_{ji}^*=x_kC_{kj}C_{ji}^*
\ee
from which
\be
 CC^*=1
 \label{quadrato}
\ee
From the product property applied to the equality
\be
R_i\ket x=\ket x x_i
\ee
we get
\be
(R_i\ket x)^*=\bra{x^*} R_i^\dagger=\bra x
CR_i^\dagger=(\ket x x_i)^*=x_i^*\bra{x^*}=x_i^*\bra x C
\ee
and therefore
\be
\bra x CR_i^\dagger C^{-1}=x_jC_{ji}\bra x=
\bra x L_j C_{ji}
\ee
that is
\be
CR_i^\dagger C^{-1}=L_jC_{ji}
\ee
or also
\be
CR_{x_i}^\dagger C^{-1}=L_{x_i^*}
\ee
If $R_i$ and $L_i$ are $^*$-representations, that is
\be
R_{x_i}^\dagger =R_{x_i^*}=R_{x_j}C_{ji}
\ee
we obtain
\be
CR_{x_i}^\dagger C^{-1}=CR_{x_i^*}C^{-1}=L_{x_i^*}
\label{sopra}
\ee
Since the involution is non-singular, we get
\be
CR_iC^{-1}=L_i
\ee
and comparing  with the adjoint of eq. (\ref{sopra}), we
see that $C$ is a unitary matrix which, from eq. (\ref{quadrato}), implies
$C^T=C$. Therefore we have the theorem:
 \\\\ {\it  Given
an associative algebra with involution, if the right and
left multiplications are $^*$-representations, then the
algebra is self-conjugated.}\\\\ In this case our integration
is a {\it state}  in the Connes terminology \cite{connes}.

If the $C$ matrix is an involution we can write the integration as
\be
 \int_{(x)} \ket x\bra{x^*}=\int_{(x)} \ket {x^*} \bra x=1
 \ee

\resection{The algebra of matrices}

Since an associative algebra admits always a matrix
representation, it is interesting to consider the definition
of the integral over the algebra $\AAA_N$ of the $N\times N$
matrices. These can be expanded in the following general way
\be
A=\sum_{n,m=1}^N e^{(nm)} a_{nm}
\ee
where $e^{(nm)}$ are $N^2$ matrices defined by
\be
e^{(nm)}_{ij}=\delta_i^n\delta_j^m,~~~~i.j=1,\cdots,N
\label{e-matrices}
\ee
These special matrices satisfy the algebra
\be
e^{(nm)}e^{(pq)}=\delta_{mp}e^{(nq)}
\label{matrix-algebra}
\ee
Therefore the structure constants of the algebra are
given by
\be
f_{(nm)(pq)(rs)}=\delta_{mp}\delta_{nr}\delta_{qs}
\ee
Recalling the definitions given in eq. (\ref{matrici}),
we have
\be
(R_{(pq)})_{(nm)(rs)}=\delta_{pm}\delta_{qs}\delta_{nr},~~~~
(L_{(pq)})_{(nm)(rs)}=\delta_{pr}\delta_{qn}\delta_{ms}
\ee
The matrix $C$ can be found by requiring that $\bra {xC}$ is
an eigenstate of $R_i$, that is
\be
[F(e)]^{(nm)}(R_{(pq)})_{(nm)(rs)}=e^{(pq)}[F(e)]^{(rs)}
\ee
where
\be
F(e)^{(nm)}=e^{(rs)}C_{(rs)(nm)}
\ee
We get
\be
[F(e)]^{(rp)}\delta_{qs}=e^{(pq)}[F(e)]^{(rs)}
\ee
By looking at the eq. (\ref{matrix-algebra}), we see
that this equation is satisfied by
\be
[F(e)]^{(rs)}=e^{(sr)}
\ee
It follows
\be
C_{(mn)(rs)}=\delta_{ms}\delta_{nr}
\ee
It is seen easily that $C$ satisfies
\be
C^T=C^*=C, ~~~~C^2=1
\ee
Therefore the matrix algebra is a self-conjugated one. One
easily checks that the right multiplications satisfy  eq.
(\ref{sopra}), and therefore $C$ is an involution. More
precisely, since
\be
{e^{(mn)}}^*=e^{(pq)}C_{(pq)(mn)}=e^{(nm)}
\ee
the involution is nothing but the hermitian conjugation
\be
A^*=A^\dagger, ~~A\in \AAA_N
\ee
The integration rules give
\be
(C^{-1})_{(rp)(qs)}=\delta_{rs}\delta_{pq}=
\int_{(e)}e^{(rp)}e^{(qs)}=\delta_{pq}\int_{(e)}e^{(rs)}
\ee
We see that this is satisfied by
\be
\int_{(e)}e^{(rs)}=\delta_{rs}
\label{int-matrici}
\ee
This result can be obtained also using directly eq.
(\ref{2.20}), noticing that the identity of the algebra is
given by $I=\sum_n e^{(n,n)}$. Therefore
\be
\int_{(e)}e^{(rs)}=\sum_n(C^{-1})_{(nn)(rs)}=\sum_n
\delta_{ns}\delta_{nr}=
\delta_{rs}
\ee
and, for a generic matrix
\be
\int_{(e)}A=\sum_{m,n=1}^Na_{nm}\int_{(e)}e^{(nm)}=Tr(A)
\ee

\resection{Derivations}

We will discuss now the derivations on associative algebras with
identity. Recall that a derivation is a linear mapping on the
algebra satisfying
\be
D(ab)=(Da)b+a(Db),~~~~~a,b\in\AAA
\label{distributivity}
\ee
We  define the action of $D$ on the basis elements in terms
of its representative matrix, $d$,
\be
Dx_i=d_{ij}x_j
\ee
If $D$ is a derivation, then
\be
S=\exp(\alpha D)
\ee
is an automorphism of the algebra. In fact, it is
easily proved that
\be
\exp(\alpha D)(ab)=(\exp(\alpha D)a)(\exp(\alpha D)b)
\label{automorphism}
\ee
On the contrary, if $S(\alpha)$ is an automorphism
depending on the continuous parameter $\alpha$, then
from (\ref{automorphism}), the following equation
defines a derivation
\be
D=\lim_{\alpha\to 0}\frac{S(\alpha)-1}\alpha
\ee
In our formalism the automorphisms play a particular
role. In fact, from eq. (\ref{automorphism}) we get
\be
S(\alpha)(\ket x x_i)=(S(\alpha)\ket x)(S(\alpha)x_i)
\ee
and therefore
\be
R_i(S(\alpha)\ket x)=S(\alpha)(R_i\ket
x)=S(\alpha)(\ket x x_i)=(S(\alpha)\ket
x)(S(\alpha)x_i)
\ee
meaning that $S(\alpha)\ket x$ is an eigenvector of
$R_i$ with eigenvalue $S(\alpha)x_i$.
This equation  shows that the basis $x_i'=S(\alpha)x_i$
satisfies an algebra with  the same structure constants
as those of the  basis $x_i$. Therefore the
matrices $R_i$ and $L_i$ constructed in the two basis, and as
a consequence the $C$ matrix,
are identical. In other words, our formulation is
invariant under automorphisms of the algebra (of course
this is not true for a generic change of basis).
The previous equation can be rewritten in terms of the matrix
$s(\alpha)$ of the automorphism $S(\alpha)$, as
\be
R_i\left(s(\alpha)\ket x\right)=\left(s(\alpha)\ket x\right)
s_{ij}x_j=s_{ij}s(\alpha)R_j\ket x
\ee
or
\be
s(\alpha)^{-1}R_i s(\alpha)=R_{S(\alpha)x}
\label{auto}
\ee
If the algebra has an identity element, $I$, (say
$x_0=I$), then
\be
Dx_0=0
\label{di}
\ee
and  therefore
\be
Dx_0=d_{0i}x_i=0 \Longrightarrow d_{0i}=0
\ee
We will prove now some properties of the derivations. First
of all, from the basic defining equation
(\ref{distributivity}) we get
\bea
R_id\ket x&=&R_i D\ket x=D(R_i\ket x=D(\ket x x_i)
\nn\\&=&d\ket x x_i+\ket x
Dx_i=
dR_i\ket x+ R_{Dx_i}\ket x
\eea
from which
\be
[R_i,d\,]=R_{Dx_i}
\label{derivata}
\ee
which is nothing but the infinitesimal version of eq. (\ref{auto}).
From the integration rules (for an algebra with
identity) we get immediately
\be
\int_{(x)}Dx_i=d_{ij}\int_{(x)}x_j=d_{ij}(C^{-1})_{0j}
\ee
Showing that in order that the derivation  $D$
satisfies the integration by parts rule for any
function, $f(x)$, on the algebra
\be
\int_{(x)}D(f(x))=0
\label{integr_part}
\ee
the necessary and sufficient condition is
\be
d_{ij}(C^{-1})_{0j}=0
\ee
implying that the $d$ matrix must be singular and have
$(C^{-1})_{j0}$ as a null eigenvector.

Next we show that, if a derivation satisfies the integration
by part formula (\ref{integr_part}), then the matrix of
related automorphism  $S(\alpha)=\exp(\alpha D)$ obeys the equation
\be
C^{-1}s^T(\alpha)C=s^{-1}(\alpha)
\label{invariance}
\ee
and it leaves invariant the measure of integration. The
converse of this theorem is also true. Let us start assuming that
$D$ satisfies eq. (\ref{integr_part}), then
\bea
0&=&\int_{(x)}D(\ket x \bra x C)=\int_{(x)}d\ket x\bra x C+\int_{(x)}
\ket x\bra {Dx} C\nn\\&=&d+\int_{(x)}\ket x \bra x CC^{-1}d^T C
=d+C^{-1}d^TC
\label{4.18}
\eea
that is
\be
d+C^{-1}d^TC=0
\label{inv-infin}
\ee
The previous expression can be exponentiated getting
\be
C^{-1}\exp(\alpha d^T)C=\exp(-\alpha d)
\ee
from which the equation (\ref{invariance}) follows, for
$s(\alpha)=\exp(\alpha d)$. To show the invariance of
the measure, let us consider the following identity
\be
1=\int_{(x)}s\ket x\bra{ x C}s^{-1}=
\int_{(x)}s\ket {x}\bra{ x }s^T C=\int_{(x)}\ket {Sx}
\bra{ Sx}C=\int_{(x)}\ket {x'}
\bra{ x'C}
\label{5.22}
\ee
where $x'=Sx$, and we have used eq. (\ref{invariance}). For
any automorphism of the algebra we have
\be
\int_{(x')}\ket {x'}\bra{ x'C}=1
\label{5.23}
\ee
since the numerical values of the matrices $R_i$ and $L_i$,
and consequently the $C$ matrix, are left invariant. Comparing
eqs. (\ref{5.22}) and (\ref{5.23}) we get
\be
\int_{(x')}=\int_{(x)}
\ee
On the contrary, if the measure is invariant under an
automorphism of the algebra, the chain of equalities
\be
1=\int_{(x')}\ket {x'}\bra{x'C}=\int_{(x)}\ket
{x'}\bra{x'C}=\int_{(x)}s\ket {x}\bra{x}C(C^{-1}s^TC)=
s(C^{-1}s^TC)
\ee
implies eq. (\ref{invariance}), together with its
infinitesimal version eq. (\ref{inv-infin}). From this (see
the derivation in (\ref{4.18})), we get
\be
0=\int_{(x)}D(x_ix_jC_{jk})
\ee
and by taking $x_i=I$,
\be
\int_{(x)}Dx_j=0
\ee
for any basis element of the algebra. Therefore we  have
proven the following theorem:\\\\ {\it If a derivation $D$
satisfies the integration by part rule, eq.
(\ref{integr_part}), the integration is  invariant under the
related automorphism $\exp{(\alpha D)}$. On the contrary, if
the integration is invariant under a continuous
automorphism, $\exp{(\alpha D)}$, the related derivation,
$D$, satisfies (\ref{integr_part}).}\\\\ This theorem
generalizes the classical result about the Riemann integral
relating the invariance under translations of
the measure and the integration by parts formula.

Next we will show that, always in the case of  an
associative self-conjugated algebra, $\AAA$, with identity,
there exists  a set of automorphisms  such that the measure
of integration is invariant. These are
 the so called
{\bf inner derivations}, that is derivations such that
\be
D\in\LL(\AAA)
\ee
where $\LL(\AAA)$ is the {\bf Lie multiplication
algebra} associated to $\AAA$. To define $\LL(\AAA)$
one starts with the linear space of left and right
multiplications and defines
\be
\MM_1=\MM_R+\MM_{L^T}
\ee
that is the space generated by the vectors
\be
R_a+L_b^T,~~~~a,b\in\AAA
\label{generators}
\ee
Then
\be
\LL(\AAA)=\sum_{i=1}^\infty\MM_i
\ee
where the spaces $\MM_i$ are defined by induction
\be
\MM_{i+1}=[\MM_1,\MM_i]
\ee
Therefore $\LL(\AAA)$ is defined in terms of all the
multiple commutators of the elements given in
(\ref{generators}).

It is not difficult to prove that for a Lie algebra,
$\LL(\AAA)$ coincides with the adjoint representation
\cite{schafer}. We will prove now an analogous result for
associative algebras with identity. That is that $\LL(\AAA)$
coincides with the adjoint representation of the Lie algebra
associated to $\AAA$ (the Lie algebra generated by
$[a,b]=ab-ba$, for $a,~b\in\AAA$). The proof can be found,
for example, in ref. \cite{schafer}, but for completeness we
will repeat it here. From the associativity conditions
(\ref{associativity}), and (\ref{associativity2}) one gets
\be
[R_a+L_b^T,R_c+L_d^T]\in\MM_1, ~~~a,b,c,d\in\AAA
\ee
or
\be
[\MM_1,\MM_1]\subset\MM_1
\ee
showing that
\be
\LL(\AAA)=\MM_1=\MM_R+\MM_{L^T}
\ee
Therefore the matrix associated to an inner derivation of an
associative algebra must be of the form
\be
d=R_a+L_b^T
\ee
We have now to require that this  indeed a derivation, that
is that  eq. (\ref{derivata}) holds. We start evaluating
\be
[R_c,d]=[R_c,R_a+L_b^T]= R_{[c,a]}
\ee
where we have used the fact that the right multiplications
form a representation of the algebra and that right and left
multiplications commute. Then comparing with
\be
R_{Dc}=R_{ca+cb}
\ee
we see that the two agree for $b=-a$.
Then we get
\be
Dx_i=x_ia-ax_i=-[a, x_i] =
-(adj\; a)_{ij}x_j
\label{commutatore}
\ee
This shows indeed that the inner derivations span the adjoint
representation of  the Lie algebra
associated to $\AAA$.

We can now proof the following theorem:\\\\{\it For an
associative self-conjugated algebra with identity, the
measure of integration is invariant  under the automorphisms
generated by the inner derivations, or, equivalently, the
inner derivations satisfy the rule of integration by
parts.}\\\\ In fact, this follows because the inner derivations satisfy eq.
(\ref{inv-infin})
\be
C^{-1}d^T
C=C^{-1}(R_a^T-L_a)C=(C^TR_a{C^T}^{-1})^T-R_a=L_a^T-R_a=-d
\ee
As an example let us consider the algebra of the matrices
studied in the previous Section. In this case the inner
derivations are simply given by
\be
D_B A=[A,B]
\ee
Therefore
\be
\int_{(e)}D_B A=\int_{(e)}[A,B]=0
\ee
and we see that the integration by parts formula corresponds to the cyclic
property of the trace.

\resection{Paragrassmann algebras as subalgebras of an
algebra of matrices}

Since an associative algebra can be  represented in terms of
matrices, and having shown that, in this case, the integration is simply
given by the trace, one can ask if it is possible to use
this result in order to get the integration over a
subalgebra of $\AAA_N$. The idea is simply that one should
integrate with the trace formula, but using a weight which
selects the particular subalgebra one is interested to. We
will illustrate this procedure for a paragrassmann algebra
of order $p$, that is an algebra generated by an element
$\theta$, such that
\be
\theta^{p+1}=0
\label{nilpotence}
\ee
For $p=1$ we get a Grassmann algebra. Any element of the
algebra is given by a power of $\theta$
\be
x_k=\theta^k, ~~~~k=0,1,\cdots, p
\ee
Being the algebra an associative  one, the elements
$\theta^k$ can be represented in terms of the right
multiplication matrices, $R_k$. These are $(p+1)\times(p+1)$
matrices given by (see \cite{integrale})
\be
(R_i)_{jk}=\delta_{i+j,k}
\ee
Defining
\be
X_\theta\equiv R_1
\ee
we can write, in terms of the matrices defined in eq.
(\ref{e-matrices})
\be
X_\theta=\sum_{i=1}^p e^{(i,i+1)}
\ee
and
\be
X_\theta^k=\sum_{i=1}^{p+1-k}e^{(i,i+k)}
\ee
Therefore, the most general function on  the paragrassmann
algebra (as a subalgebra of the matrices $(p+1)\times
(p+1)$) is given by
\be
f(X_\theta)=\sum_{i=1}^{p+1}a_iX_\theta^{p+1-i}=
\sum_{i=1}^{p+1}a_i\sum_{j=1}
^i
e^{(j,p+1+j-i)}
\ee
As we said, the idea is to look for  a matrix $P$ such that
it projects out of the algebra, $\AAA_{p+1}$ of the
$(p+1)\times  (p+1)$ matrices, the paragrassmann subalgebra.
To define such an operator, let us consider a generic matrix
$B\in \AAA_{p+1}$. We can always decompose it as (see later)
\be
B=f(X_\theta)+\tilde B
\label{decomposition}
\ee
The operator $P$ should satisfy
\be
BP=f(X_\theta)P
\ee
or
\be
\tilde B P=0
\ee
Then, we can define the integration over the
paragrassmann algebra in terms of the integration over
the isomorphic subalgebra of $\AAA_{p+1}$ through the
equation
\be
\int_{(\theta)}f(\theta)=\int_{(e)} f(X_\theta)P=
Tr[f(X_\theta)P]
\ee
In order to define the decomposition (\ref{decomposition})
and the operator $P$, let us consider the most general
$(p+1)\times (p+1)$ matrix. We can write
\be
B=\sum_{i,j=1}^{p+1} b_{ij}e^{(ij)}=\sum_{i=1}^{p+1}\sum_{j=1}^p
b_{ij}e^{(ij)}+\sum_{i=1}^{p+1}b_{i,p+1}e^{(i,p+1)}
\label{B-matrix}
\ee
By adding and subtracting
\be
\sum_{i=2}^{p+1}b_{i,p+1}\sum_{j=1}^{i-1}e^{(j,p+1+j-i)}
\ee
we get the decomposition (\ref{decomposition}) with
\be
f(X_\theta)=\sum_{i=1}^{p+1}b_{i,p+1}X_\theta^{p+1-i}
\label{effe-e}
\ee
and
\be
\tilde B=\sum_{i=1}^{p+1}\sum_{j=1}^pb_{ij}e^{(ij)}-\sum_{i=2}^{p+1}
b_{i,p+1}\sum_{j=1}^{i-1}e^{(j,p+1+j-i)}
\ee
Let us notice that for any integer $k$, $1\le k\le p+1$, we have
\be
\tilde Be^{(p+1,k)}=0
\ee
or
\be
Be^{(p+1,k)}=f(X_\theta)e^{(p+1,k)}
\ee
But using the identity
\be
e^{(p+1,k)}=e^{(p+1,1)}X_\theta^{k-1}
\ee
we get
\be
Tr[Be^{(p+1,k)}]=Tr[X_\theta^{k-1}f(X_\theta)e^{(p+1,1)}]\equiv
Tr[g(X_\theta)e^{(p+1,1)}]
\ee
where $g(x_\theta)=X^{k-1}f(X_\theta)$. This shows that we can
always define the integration trough the operator
$P=e^{(p+1,1)}$. Then, the integration over the
subalgebra is given by
\be
\int_{\theta}f(\theta)=\int_{(e)}f(X_\theta)e^{(p+1,1)
}=
Tr[f(X_\theta)e^{(p+1,1)}]
\ee
It follows from eq. (\ref{effe-e}), and eq. (\ref{int-matrici})
\be
\int_{(\theta)}f(\theta)=Tr[\sum_{i=1}^{p+1}b_{i,p+1}e^{(i,1)}]
=b_{1,p+1}
\ee
Meaning that
\be
\int_{(\theta)}f(\theta)=\int_{(\theta)}[b_{1,p+1}\theta^p+
b_{2,p+1}\theta^{p-1}+\cdots+b_{p+1,p+1}]=b_{1,p+1}
\ee
or
\be
\int_{(\theta)}\theta^k=\delta_{kp}
\label{6.23}
\ee
which coincides with the direct way of defining the integral
over a paragrassmann algebra (see \cite{integrale}). Of
course, by choosing $P=e^{(p+1,k)}$ with $k\not=1$ would
lead to an integral, which can be expressed in terms of the
one in eq. (\ref{6.23}), as
\be
\int_{(\theta)}\theta^{k-1}f(\theta)
\ee
In the particular case of a Grassmann algebra we have
\be
X_\theta=e^{(1,2)}=\left(\matrix{0 & 1\cr 0 & 0}\right)=\sigma_+,~~~
P=e^{(2,1)}=\left(\matrix{0 & 0\cr 1 & 0}\right)=\sigma_-
\ee
The decomposition in eq. (\ref{decomposition}), for a
$2\times 2$ matrix
\be
B=a+b\sigma_3+c\sigma_++d\sigma_-
\ee
is given by
\be
\tilde B=b(1+\sigma_3)+d\sigma_-,~~~~f(X_\theta)=f(\sigma_+)=
a-b+c\sigma_+
\ee
and the integration is
\be
\int_{(\theta)}f(\theta)=Tr[f(\sigma_+)\sigma_-]
\ee
from which
\be
\int_{(\theta)} 1=Tr[\sigma_-]=0,~~~~
\int_{(\theta)} \theta=Tr[\sigma_+\sigma_-]=1
\ee
We notice that the matrices $\tilde B$ and $f(X_\theta)$
appearing in the decomposition (\ref{decomposition}) can be
written more explicitly as
\be
\tilde B=\left(\matrix{
\tilde b_{1,1} & \tilde b_{1,2} & \cdots & \tilde b_{1,p} & 0\cr
\cdot         &   \cdot       &   \cdot &    \cdot     &   \cdot\cr
\cdot      &   \cdot        &       \cdot &       \cdot &     \cdot\cr
\tilde b_{p,1} & \tilde b_{p,2} & \cdots & \tilde b_{p,p} & 0\cr
\tilde b_{p+1,1} & \tilde b_{p+1,2} & \cdots & \tilde b_{p+1,p} & 0\cr}
\right)
\ee
and
\be
f(X_\theta)=\left(\matrix{ a_{p+1} & a_p & a_{p-1} & \cdots  &a_2
&a_1\cr 0 & a_{p+1} & a_p & \cdots & a_3 & a_2\cr 0 & 0 & a_{p+1}
& \cdots & a_4 & a_3\cr
\cdot & \cdot & \cdot & \cdot & \cdot & \cdot\cr
\cdot & \cdot & \cdot & \cdot & \cdot & \cdot\cr
 0 & 0 & 0 & \cdots & a_{p} & a_{p-1} \cr
 0 & 0 & 0 & \cdots & a_{p+1} & a_p \cr
 0 & 0 & 0 & \cdots & 0 &
a_{p+1}\cr}\right)
\ee
The $p\times (p+1)$ parameters appearing in $\tilde B$ and
the $p+1$ parameters in $f(X_\theta)$ can be easily
expressed in terms of the $(p+1)\times (p+1)$ parameters
defining the matrix $B$.

\resection{Conclusions}

In this paper we have studied some general properties of the
integration over self-conjugated associative algebra with
identity.  We have proven a theorem showing that continuous
automorphisms, leaving invariant the measure of integration,
give rise to derivations satisfying the integration by part
rule (that is the vanishing of the integral of a
derivative). The relevance of this result is that in quantum mechanics the
Schwinger's principle
follows trivially from the previous theorem, therefore it opens
the  avenue to extensions to more general
theories.

The other important problem we have considered is the
following: given the integral over an algebra, is there a
natural way to induce the integral over a subalgebra? To
face this problem we have followed the way suggested by the
standard integration over manifolds, that is to project out
(via the characteristic function) the submanifold in the
measure. We have illustrated this procedure for a
paragrassmann algebra of order $p$, thought as a subalgebra
of an algebra of $(p+1)\times(p+1)$ matrices, showing that
it is possible define a projector selecting the
paragrassmann component out of a given matrix. Since in the
text we have shown that the integral for an algebra of
matrices coincides with the trace, we arrive to evaluate the
integral over a paragrassmann algebra via a trace of
ordinary matrices. In particular the integration over the
one-dimensional Grassmann algebra can be expressed as a
trace of $2\times 2$ matrices. This result might be of some
interest for fermionic theories on the lattice.

\medskip
\begin{center}
{\bf Acknowledgements}
\end{center}
\medskip
The author would like
 Prof. J. P. Eckmann, Director of the Department of Theoretical
  Physics of the University
of Geneva, for the very kind hospitality.


\end{document}